\documentclass[twocolumn,showpacs,showkeys,preprintnumbers,amsmath,amssymb,prd]{revtex4}

\usepackage{graphicx}% Include figure files
\usepackage{dcolumn}% Align table columns on decimal point
\usepackage{bm}% bold math
\begin{document}

\title{Three-cocycles, Nonassociative  Gauge Transformations and Dirac's Monopole}
\author{Alexander I. Nesterov}
   \email{nesterov@cencar.udg.mx}
\affiliation{Departamento de F{\'\i}sica, CUCEI, Universidad de Guadalajara,
Av. Revoluci\'on 1500, Guadalajara, CP 44420, Jalisco, M\'exico}

\date{}

\begin{abstract}
The relation between 3-cocycles arising in the Dirac monopole problem and
nonassociative gauge transformations is studied. It is  shown that
nonassociative extension of the group U(1) allows to obtain a consistent
theory of pointlike magnetic monopole  with an arbitrary magnetic charge.
\end{abstract}


                              %display desired

\maketitle

\section{Introduction}

In 1931 Dirac \cite{Dir} introduced a magnetic monopole into the
quantum mechanics and found a quantization relation between an
electric charge $e$ and magnetic charge $q$, $2\mu =n, \; n \in
\mathbb Z$, where $\mu =eq$, and $\hbar = c=1$.  One of the widely
accepted proofs of the Dirac  selection rule is based on group
representation theory (see, for example,
\cite{Gol1,Gol2,Zw_1,H,LWP}).

In the presence of the magnetic monopole the operator of the total angular
momentum ${\mathbf J}$, which uncludes contribution of the electromagnetic
field, obeys the standard commutation relations of the Lie algebra of the
rotation group
\[
[J_i, J_j] = i\epsilon_{ijk}J_k.
\]
and this is true for any value of $\mu$. Notice that these commutation
relations fail on the Dirac string, restricting the domain of definition of the
operator $\mathbf J$ and limiting it to the functions that vanish sufficiently
rapidly on the string \cite{Zw_1}. Here $\mathbf J$ may be extended to
self-adjoint operator satisfying the commutation relations of the rotation
group for an arbitrary $\mu$ \cite{H}.

The requirement that $J_i$ generate a finite-dimensional representation of the
rotation group yields $2\mu$ being integer and only values $2\mu = 0,
\pm1,\pm2, \dots$ are allowed. However, employing infinite-dimensional
representations of the rotation group one can relax Dirac's condition and
obtain the consistent monopole theory with an arbitrary magnetic
charge\cite{NF,NF2,NF1}.

The situation with the translations in the background of the monopole is
completelely different from the rotations. Since the Jacobi identity fails for
the gauge  invariant algebra of translations they  do not form the Lie algebra,
and for the finite translations one has \cite{Jac,Gr,G1,G2}
\begin{align}
\bigl(U_{\mathbf a}U_{\mathbf b}\bigr)U_{\mathbf c}\Psi(\mathbf r) ={\rm
e}^{i\alpha_3(\mathbf r ;\mathbf a, \mathbf b,\mathbf c)} U_{\mathbf
a}\bigl(U_{\mathbf b}U_{\mathbf c}\bigr)\Psi(\mathbf r) \label{as}
\end{align}
where $\alpha_3$ is the so-called 3-cocycle. For the Dirac quantization
condition being satisfied, one has $\alpha_3 = 0 \mod (2\pi n)$; and
Eq.(\ref{as}) provides an associative representation of the translations, in
spite of the fact that the Jacobi identity for the infinitesimal generators
continues to fail. The 3-cocycles, besides being appeared in the Dirac monopole
background, lead also to a nonassociative algebra of defects in
quantum-mechanical systems and are related to axions in string theory
\cite{Gr1,Com}. The low-dimensional cocycles are known also in modern physics
in relation with the non-Abelian anomalies and Wess-Zumino action. For
instance, the Wess-Zumino functional is identified as 1-cocycle and the
anomalous Schwinger term as a 2-cocycle \cite{WZ,LF,FS}.

In this letter we introduce nonassociative gauge transformations and give
a new interpretation of 3-cocycle linking it with the associator of the
gauge loop. We argue that a nonassociative extension of the conventional group U(1)
allows an arbitrary magnetic charge.

\section{Cocycles and magnetic monopole}

Following \cite{LF} consider a group $G$ acting on a manifold $\mathfrak
M$ as a group of transformations: $x\rightarrow {x}g, \; x\in \mathfrak M,
\; g\in G$. For an array function $\alpha_n({x};g_1,\dots, g_n)$, called
the {\em n-cochain}, the coboundary operator $\delta$ is defined as
\begin{align}\label{d}
&(\delta\alpha_n)({{x};g_1,g_2,\dots, g_{n+1}})=
\alpha_n({x}g_1;g_2,\dots, g_{n+1}) \nonumber \\
&-  \alpha_n({x};g_1g_2,\dots, g_{n+1}) + \dots \nonumber \\
&+(-1)^i \alpha_n({x};g_1,\dots,g_ig_{i+1} ,\dots,g_{n+1})\nonumber \\
&+(-1)^{n+1}\alpha_n({x};g_1,\dots,g_n),
\end{align}
and it is easy to check that $\delta^2 =0$. A cochain $\alpha_n=
\delta\alpha_n$ is called a {\em coboundary,} and if $\alpha_n$ satisfies
$\delta \alpha_n = 0$, it is called a {\em cocycle}. A cocycle is
nontrivial if it is not a coboundary, i.e. $\delta \alpha = 0$, but
$\alpha \neq \delta\beta$. Notice, that the $n$-cochain can be associated
with a differential $n$-form $\boldsymbol\omega_n$ as
\begin{equation}\label{d1}
    \alpha_n({{x};g_1,g_2,\dots, g_{n}})= \int_{\Sigma_n}\boldsymbol
    \omega_n,
\end{equation}
where $\Sigma_n$ is a $n$-simplex with vertices $(x, xg_1, \dots,
xg_1\dots g_n)$.

The 1-cochain $\alpha_1(x; g)$ appears in the representation $U_g$ of the
group $G$ in the space of the functions $\Psi(x)$ on $\mathfrak M$:
\begin{equation}\label{R}
   U_{g}\Psi(x)={\rm e}^{i\alpha_1(x; g)}\Psi(xg).
\end{equation}
If $\alpha_1(x; g)$ is a trivial cocycle,
\begin{equation}\label{4_1}
\alpha_1(x; g)= \delta\alpha_0(x; g)=\alpha_0(xg)-\alpha_0(x),
\end{equation}
then the representation reduces to the usual one
\begin{equation}\label{R1}
   U_{g_1}U_{g_2}\Psi(x)=U_{g_1g_2}\Psi(x),
\end{equation}
by the unitary transformation $\Psi(x)\mapsto \exp(i\alpha_0(x))\Psi(x)$.

A 2-cocycle $\alpha_2 = \delta\alpha_1$ defined for a given 1-cochain
$\alpha_1$ by
\begin{align}\label{4_2}
  \alpha_2(x; g_1,g_2)= \alpha_1(xg_1;g_2)-\alpha_1(x;g_1g_2)
  +\alpha_1(x;g_1)
\end{align}
appears in the product of two operators:
\begin{equation}\label{R2}
   U_{g_1}U_{g_2}\Psi(x)={\rm e}^{i\alpha_2(x; g_1,g_2)}U_{g_1g_2}\Psi(x).
\end{equation}
When the 2-cocycle does not depend on $x$, Eq. (\ref{R2}) provides a {\em
projective representation} of the group $G$, i.e.
\begin{equation}\label{R3}
   U_{g_1}U_{g_2}={\rm e}^{i\alpha_2(g_1,g_2)}U_{g_1g_2}.
\end{equation}
The condition of associativity for (\ref{R3}) yields $\alpha_3
=\delta\alpha_2=0$, and for $\alpha_2$ being a trivial 2-cocycle the
representation (\ref{R3}) reduces to an ordinary one.

Let $\alpha_2(x;g_1,g_2)$ be a 2-cochain, then  the 3-cocycle
$\alpha_3=\delta\alpha_2$, is given by
\begin{align}\label{4_7}
\alpha_3(x; g_1,g_2,g_3)=& \alpha_2(xg_1; g_2,g_3)-\alpha_2(x;g_1g_2,g_3)+\nonumber\\
&+  \alpha_2(x;g_1,g_2g_3) -\alpha_2(x;g_1,g_2).
\end{align}
The 3-cocycle enters in the triple-product in the similar way as
$\alpha_2$ appears in the product of two operators:
\begin{equation}\label{4_8}
(U_{g_1}U_{g_2})U_{g_3}={\rm
e}^{i\alpha_3(x;g_1,g_2,g_3)}U_{g_1}(U_{g_2}U_{g_3}).
\end{equation}
Associativity is fulfilled if $\alpha_3 = 0 \mod (2\pi n)$, $n\in \mathbb
Z$, and we see that the 3-cocycle measures the lack of associativity.

In what follows we consider a generalization of the translation group
related with the cocycles and {\em string group} \cite{JM}. The string
group, denoted as String$\,\mathfrak M$, is the group of all paths
$\gamma$: $[0,1] \mapsto$ Diff$\,\mathfrak M$, where Diff$\,\mathfrak M$
denotes the diffeomorphism group on $\mathfrak M = R^3\backslash\{0\}$
such that $ {\mathbf r} \mapsto {\mathbf r}(t)= {\mathbf r}\gamma(t), \;
t\in[0,1] $ and $\gamma(0)=$ identity; the group composition is defined as
$\gamma_{12}(t) = \gamma_1(t)\gamma_2(t)$.

In the presence of magnetic monopole we define the 1-cochain $\alpha_1$ as
\begin{align}
\alpha_1(\mathbf r;\gamma)=  e \int_{\mathbf r}^{{\mathbf r}'} \mathbf
A(\boldsymbol \xi) \cdot d \boldsymbol \xi \label{g_1c}
\end{align}
where the integration is performed along a path $\gamma$ connecting a
point $\mathbf r$ with a point ${\mathbf r}' = \mathbf r\gamma(1)$, and
$\mathbf A(\mathbf r)$ is the vector potential. If the integral does not
depend on the path, one has
\[
\alpha_1 = \delta\alpha_0 = \alpha_0(\mathbf r) - \alpha_0({\mathbf r}'),
\]
and $\alpha_1$ is the trivial 1-cocycle.

Following \cite{Jac,Gr,G2,JM} we introduce
\begin{equation}
\alpha_{2}(\mathbf r;\gamma_{1},\gamma_{2})= e \int_{\Sigma} \mathbf
B \cdot d{\mathbf S} = e\Phi\big|_\Sigma \label{g_2c}
\end{equation}
where $\Phi\big|_\Sigma$ is a magnetic flux through the two-dimensional
simplex $\Sigma$ defined as follows: The surface is parametrized as
$\mathbf r(t,s)=\mathbf r\gamma_1(t) \gamma_2(s)$ with  $0\leq s\leq t
\leq 1$, and the vertices are $(\mathbf r,{{\mathbf r}}_1,{{\mathbf
r}}_2)$, where ${{\mathbf r}}_1 =\mathbf r\gamma_1(1)$ and ${{\mathbf
r}}_2 =\mathbf r\gamma_2(1)$.

Notice that any choice of the vector potential $\mathbf A$ being
compatible with a magnetic field ${\mathbf B}= q {\mathbf r}/r^3$ of Dirac
monopole must have  singularity (the so-called  Dirac string). This implies
$\mathbf B = \nabla \times \mathbf A$ locally, but not globally, so that
$\alpha_2$ is a 2-cochain and not a 2-cocycle. Thus, one can write
\begin{equation}{\mathbf
B}={\rm rot}{\mathbf A} + {\mathbf h}
\label{A_4}
\end{equation}
where ${\mathbf h}$ is the magnetic field of the string $\mathcal C$ .
Substituting (\ref{A_4}) into (\ref{g_2c}) and applying Stokes' theorem we
get
\begin{equation}
\alpha_{2}(\mathbf r;\gamma_{1},\gamma_{2}) = \delta\alpha_{1}(\mathbf
r;\gamma_{1},\gamma_{2}) + \sigma (\mathcal C, \Sigma) \label{A6}
\end{equation}
where
\begin{equation}
\delta\alpha_1=\alpha_1({{\mathbf r}}_1;\gamma_{2})- \alpha_1(\mathbf
r;\gamma_1 \gamma_{2})+ \alpha_1(\mathbf r ;\gamma_{1}),
\end{equation}
and the contribution $\sigma =e\int_{\Sigma}{\mathbf h}\cdot d\mathbf S$
is not zero if and only if the string $\mathcal C$ crosses $\Sigma$.

Computation of the 3-cocycle,
\begin{align}
&\alpha_3(\mathbf r;\gamma_{1},\gamma_{2},\gamma_{3})= \delta\alpha_2=
\alpha_2({{\mathbf r}}_1;\gamma_{2},\gamma_{3})
-\alpha_2(\mathbf r;\gamma_{1}\gamma_{2},\gamma_{3}) \nonumber\\
&+\alpha_2(\mathbf r;\gamma_{1},\gamma_{2}\gamma_{3}) -\alpha_2(\mathbf
r;\gamma_{1},\gamma_{2}), \label{g_2e}
\end{align}
yields $\alpha_3= 4\pi \mu\; \mod (2\pi n)$ if the monopole is
enclosed by the simplex with vertices being $(\mathbf r,{{\mathbf r}}_1,
{{\mathbf r}}_2,{{\mathbf r}}_3)$ or zero otherwise.

The cocycles allow to construct the realization of the group String $\mathfrak
M$ in the space of complex functions on $\mathfrak M$ in the following way. Let
$\mathcal G$ be the gauge group of maps $\lambda_\gamma$: $\mathfrak M \mapsto
$ U(1), $\gamma \in{\rm String}\,\mathfrak M $. Then the extension
String${^g}\;\mathfrak M$ is defined as the product of String$\,\mathfrak M
\otimes \,\mathfrak M$ with nonassociative multiplication \cite{JM}
\begin{equation}\label{Str2}
(\gamma_1,\lambda_{\gamma_1})(\gamma_2,\lambda_{\gamma_2})=(
\gamma_1\gamma_2,{\mathcal A}_2(\mathbf r;
\gamma_1,\gamma_2)\lambda_{\gamma_{12}}),
\end{equation}
where $\lambda_{\gamma}(\mathbf r)= \lambda(\mathbf r\gamma(1))$, and
${\mathcal A}_2$ is a U(1)-valued function on $\mathfrak M$ given by
${\mathcal A}_2(\mathbf r; \gamma_1,\gamma_2)=\exp(i\alpha_{2}(\mathbf
r;\gamma_1,\gamma_2))$; $\alpha_{2}$ being the 2-cochain  of Eq. (\ref{g_2c}).

The realization of the  group String$\,\mathfrak M$ is defined as follows:
\begin{equation}\label{U1}
  U_{g_\gamma}\Psi(\mathbf r)= {\mathcal A}_1(\mathbf r; \gamma)\Psi(\mathbf r\gamma),
  \quad U_{g_\gamma}\in {\rm U(1)}
\end{equation}
where ${\mathcal A}_1(\mathbf r; \gamma)=\exp(i \alpha_1(\mathbf r;
\gamma))$, and  the composition law for the two operators ${\mathcal A}_1$
is taken to be \cite{Jac,LF}
\begin{equation}\label{U2}
{\mathcal A}_1(\mathbf r; \gamma_1){\mathcal A}_1(\mathbf r\gamma_1;
\gamma_2)= {\rm e}^{i\alpha_2(\mathbf r; \gamma_1,\gamma_2)}{\mathcal
A}_1(\mathbf r; \gamma_{12}).
\end{equation}
The 2-cochain $\alpha_2$ enters in the product of the two operators
$U_{g_\gamma}$ as follows:
\begin{align}
U_{g_{\gamma_1}}U_{g_{\gamma_2}}\Psi(\mathbf r) = {\rm e}^{i\alpha_2(\mathbf r;
\gamma_1,\gamma_2)} U_{g_{\gamma_{12}}}\Psi(\mathbf r).
\end{align}

The nonassociativity of multiplication is described by the 3-cocycle $\alpha_3$
which appears in the triple-product:
\begin{align*}
U_{g_{\gamma_1}}\big(U_{g_{\gamma_2}}U_{g_{\gamma_3}}\big)\Psi(\mathbf r)= {\rm
e}^{i\alpha_3(\mathbf r,
\gamma_1,\gamma_2,\gamma_3)}\big(U_{g_{\gamma_1}}U_{g_{\gamma_2}}\big )
U_{g_{\gamma_3}}\Psi(\mathbf r).
\end{align*}
Just as in Eq. (\ref{g_2e}) we find that $\alpha_3= 4\pi \mu$ if the monopole
is  enclosed by the simplex with vertices being $(\mathbf r,{{\mathbf r}}_1,
{{\mathbf r}}_2,{{\mathbf r}}_3)$ and zero otherwise. The obtained realization
is called a {\em nonassociative} representation of the group \cite{Jac}.
Associativivty is fulfilled if $\alpha_3 = \delta \alpha_2 = 0 \mod (2\pi n)$
and leads to the Dirac's quantization condition, $2\mu \in \mathbb Z$.

\section{Nonassociative gauge transformations}

We start with some elementary facts from the theory of quasigroups and loops
(see, e.g. \cite{Ch1,P1,S1} ).

A set $\langle Q,{\mathbf\cdot}\rangle $ with a binary operation
$(a,b) \mapsto a {\mathbf\cdot} b$ where the equations $a{\mathbf\cdot}
x=b,~y{\mathbf\cdot} a=b$ have unique solutions in $Q$ for all $a,b \in Q$, is
called a {\it quasigroup}. A {\it loop} is a quasigroup with a two-sided
identity, $a{\mathbf\cdot} e= e{\mathbf\cdot} a=a, \forall a \in Q$. A loop
$\langle Q,{\mathbf\cdot},e \rangle$ with a smooth functions
$\phi(a,b):=a{\mathbf\cdot} b$ is called a {\it smooth loop}. For $\langle
Q,{\mathbf\cdot},e\rangle$ being a  local loop with a neutral element $e$ one
has the following identity of quasiassociativity
\begin{equation}
a{\mathbf\cdot}( b{\mathbf\cdot} c) =(a{\mathbf\cdot}
b){\mathbf\cdot}l_{(a,b)} c \label{Ll}
\end{equation}
where $l_{(a,b)}$ is an {\it  associator}.

Let $\langle {\rm QU(1)}, \ast,e \rangle$ be the loop of
multiplication by unimodular complex numbers related with the group U(1)as
follows: Let $U_g = \exp({ig})$, then we define the operation $\ast$ in  QU(1) by
\begin{align}
&U_{g_1}\ast U_{g_2}= {\rm e}^{i\alpha_2({g_1,g_2})}U_{g_{12}}, \;
\label{q_4a}\\
&U_{l(g_1,g_2)g_3} ={\rm e}^{i\Delta({g_1,g_2,g_3})}U_{g_{3}} \quad U_{g_1},
U_{g_2} \in {\rm QU(1)},\label{q_4ab}
\end{align}
where $U_{g_{12}}= U_{g_1}U_{g_2}$ denotes the conventional multiplication
of the elements of the group U(1), $\alpha_2(g_1,g_2)$ is a 2-cochain; and
$\Delta({g_1,g_2,g_3})$ will be specified below. Further we assume that
the following commutation relations hold:
\begin{align}
&{\rm e}^{i\alpha_2({g_1,g_2})}U_{g} = U_{g}{\rm
e}^{i\alpha_2({g_1,g_2})}, \\
&{\rm e}^{i\Delta({g_1,g_2,g_3})}U_{g}=U_{g}{\rm
e}^{i\Delta({g_1,g_2,g_3})}.
\label{q_4abc}
\end{align}

For the product of the three elements we find
\begin{align*}
&U_{g_1}(U_{g_2} U_{g_3})= {\rm e}^{-i(\alpha_2(g_1,g_{23})+\alpha_2({g_2,g_3}))}
U_{g_1}\ast(U_{g_2}\ast U_{g_3}),\\
&(U_{g_1}U_{g_2}) U_{g_3}= {\rm
e}^{-i(\alpha_2(g_{12},g_{3})+\alpha_2({g_1,g_2}))}(U_{g_1}\ast
U_{g_2})\ast U_{g_3}.
\end{align*}
Using the associativity of multiplication in the group U(1),
\begin{equation*}
U_{g_1}(U_{g_2} U_{g_3})=(U_{g_1}U_{g_2}) U_{g_3},
\end{equation*}
we obtain
\begin{align}
U_{g_1}\ast(U_{g_2}\ast U_{g_3})= {\rm
e}^{i\alpha_3(g_1,g_2,g_3)}(U_{g_1}\ast U_{g_2})\ast U_{g_3} \label{q_4c},
\end{align}
where the 3-cocycle $\alpha_3(g_1,g_2,g_3)$ is given by
\[
\alpha_3=\alpha_2({g_2,g_3})- \alpha_2(g_{12},g_{3}) +
\alpha_2(g_1,g_{23}) -\alpha_2({g_1,g_2}).
\]

From the other hand, the product (\ref{q_4a}) can be written as
\begin{equation}\label{q4a}
U_{g_1}\ast U_{g_2} =U_{g_1{\mathbf\cdot} g_2},
\end{equation}
where $g_1{\mathbf\cdot} g_2 = g_{1} + g_2 + \alpha_2({g_1,g_2})$. Taking into
account Eqs. (\ref{Ll}), (\ref{q_4ab}) and (\ref{q_4abc}), we find that the
triple-product satisfies
\begin{align}
U_{g_1}\ast(U_{g_2}\ast U_{g_3})= {\rm
e}^{i\Delta(g_1,g_2,g_3)}(U_{g_1}\ast U_{g_2})\ast U_{g_3} \label{q4d},
\end{align}

Comparing (\ref{q_4c}) and (\ref{q4d}) we conclude that $\Delta({g_1,g_2,g_3})=
\alpha_3(g_1,g_2,g_3)$, and  thus, the associator is related to the 3-cocycle
as follows:
\begin{align}
l_{(g_1,g_2)}{g_3}=  g_3 +\alpha_3(g_1,g_2,g_3). \label{q_4d}
\end{align}
Since the loop QU(1) is isomorphic to the group U(1) if and only if the associator is
identity map, one can see that this is equivalent to the vanishing of the 3-cocycle.

Assuming QU(1) to be a local loop, further being called the {\em gauge loop},
let us consider the wave function
\[
\Psi(\mathbf r) = {\rm e}^{ig(\mathbf r)}\psi(\mathbf r), \quad{\rm e}^{ig}\in {\rm
QU(1)}
\]
and define the covariant derivative as
\begin{equation}
D_\mu  \Psi = =\bigl (\partial_\mu -i e A_\mu\bigr) \Psi
 \label{der}
\end{equation}
where $A_\mu$ is a ``nonassociative'' gauge field. Now applying (\ref{q_4a}) we
find that the wave function $\Psi$ and gauge field $\mathbf A$ transform as
\begin{align}
 &U_{\tilde g}\ast\Psi ={\rm e}^{i(\tilde g + \alpha_2(\tilde g,g))}\Psi, \label{A2}\\
&\mathbf A' =\mathbf A+ \nabla\tilde g + \nabla\alpha_2(\tilde g,g).\label{A3}
\end{align}
It is easy to check that the curvature 2-form, $F=dA$, is invariant under the
nonassociative gauge transformations.

{\em Gauge loop in Dirac's monopole problem.} - Since  the 3-cocycle
$\alpha_3(\mathbf r; g_1,g_2,g_3)$ emerges in  the Dirac monopole problem as a
general cocycle \cite{LF,FS}, i.e. it depends on the point $\mathbf r$ and the
group elements, one has to extend the construction considered above. We
generalize it in the following way: Let $\mathcal Q$ be the local loop of maps
$f_\gamma$: $\mathfrak M \mapsto $ QU(1), $\gamma \in{\rm String}\,\mathfrak M
$.  We define the extension String${^q}\;\mathfrak M$ as the product of ${\rm
String}\,\mathfrak M \otimes\,\mathcal Q $ with
\begin{equation}\label{Str2a}
(\gamma_1,f_{\gamma_1})(\gamma_2,f_{\gamma_2})=(\gamma_1\gamma_2,
f_{\gamma_1}\ast{f_{\gamma_2}})
\end{equation}
where the nonassociative product $f_{\gamma_1}\ast{f_{\gamma_2}}$ is
specified as follows: Let $f_\gamma$ be the map
\[
 f_\gamma: \mathbf r
\mapsto U_{g_\gamma} = \exp{\big(i\alpha_1(\mathbf r;\gamma)\big)}\in
\mathcal Q,
\]
then we define
\begin{align}
\label{A7}
&U_{g_{\gamma_1}}\ast
U_{g_{\gamma_2}}=U_{g_{\gamma_1}{\mathbf\cdot} {g_{\gamma_2}}} = {\rm
e}^{i\alpha_2(\mathbf r; \gamma_1,\gamma_2)}U_{g_{\gamma_{12}}},\\
&g_{\gamma_1}{\mathbf\cdot}{g_{\gamma_2}}=\alpha_1(\mathbf r; \gamma_1)+
\alpha_1({\mathbf r}_1; \gamma_2) +  \sigma (\mathcal C, \Sigma)),
\end{align}
where ${\mathbf r}_1= \mathbf r\gamma_1(1)$, and $\sigma (\mathcal C,
\Sigma)$ is the contribution of the Dirac string (see Eq.(\ref{A6})). The
2-cochain $\alpha_2$ obyes the following commutation relations:
\begin{equation}\label{A(}
U_{g_{\gamma}}{\rm e}^{i\alpha_2(\mathbf r; \gamma_1,\gamma_2)}={\rm
e}^{i\alpha_2({\mathbf r}'; \gamma_1,\gamma_2)}U_{g_{\gamma}}, \quad
{\mathbf r}'= \mathbf r\gamma(1).
\end{equation}
For the triple-product we have the same result as has been obtained above (see
Eq. (\ref{q_4c})):
\begin{equation}\label{A5a}
U_{g_{\gamma_1}}\ast (U_{g_{\gamma_2}}\ast U_{g_{\gamma_3}})= {\rm
e}^{i\alpha_3(\mathbf r; \gamma_1,\gamma_2,\gamma_3)
}(U_{g_{\gamma_1}}\ast U_{g_{\gamma_2}})\ast U_{g_{\gamma_3}} \nonumber
\end{equation}
where $\alpha_3 = 4\pi \mu$ is  the 3-cocycle.

Dirac's ideas lead naturally to consideration of a quantum mechanics in which
the wave function has a non-integrable (or path-dependent) phase factor.  This
formalism has been developed by various authors (see, for example
\cite{DeW,Bel,M,CF,R,Wu1,Wu2,ZS,E} and references therein). In general, the
path-depndent wave function can be written as
\begin{equation}\label{M1}
\Psi(\mathbf r, \gamma) = {\rm e}^{-ie\int_\gamma {\mathbf A}\cdot d
\mathbf r}\psi(\mathbf r)
\end{equation}
where the integration is performed along the path $\gamma$ joining the
reference point ${\mathbf r}_0$ with a point $\mathbf r$, and $\psi(\mathbf r)$
is a single-valued function.

Here, instead of taking the reference point for the construction of paths, we
employ the group String $\mathfrak M$ and introduce the path-dependent
nonassociative wave function as
\begin{align}
\Psi(\mathbf r, \gamma)= {\rm e}^{i\alpha_1(\mathbf r, \gamma)} \Psi(\mathbf
r), \; \Psi(\mathbf r, \gamma)\in {\mathcal Q},\; \gamma \in {\rm
String}\,{\mathfrak M}.
\end{align}
The realization of the gauge loop $\mathcal Q$ in the space of the wave
functions $\Psi(\mathbf r, \gamma)$ is given by
\begin{equation}\label{A4}
U_{g_{\gamma'}}\ast\Psi(\mathbf r; \gamma) ={\rm e}^{i(\alpha_1(\mathbf r;
{\gamma}') + \sigma (\mathcal C, \Sigma))}\Psi(\mathbf r; {\gamma}'\gamma),
\end{equation}
where $U_{g_{\gamma'}}=\exp({i\alpha_1(\mathbf r; \gamma')}\in{\mathcal Q }$.
Having nonassociative gauge transformations and wave functions, one can define
the nonassociative vector potential as above (see Eqs. (\ref{der}) -
(\ref{A3})).

\section{Concluding remarks}

Using of nonassociative structures is unavoidable when the Jacobi identity
fails. The emerging difficulties associated with the contradictory
nonassociative representations of the gauge group in the presence of the Dirac
monopole may be removed by introducing nonassociative gauge transformations,
which are related to the theory of quasigroups and loops. In quantum mechanical
description of the monopole we deal with nonassociative path-dependent wave
function $\Psi(\mathbf r; \gamma)$, where $\gamma \in {\rm String}\;\mathfrak
M$. Returning to the main subject, we see that nonassociative extension of the
conventional group U(1) allows to avoid the Dirac quantization condition and
obtain the consistent magnetic monopole theory with an arbitrary magnetic
charge.

We close with some remarks about the relevance of this investigation to quantum
mechanics and gauge field theory. Since a conventional quantum mechanics
requires a linear Hilbert space and associative operators, the Dirac
quantization rule is a necessary condition for the consistence of quantum
mechanics. Relaxation of this condition implies introducing of a {\em
nonassociative algebra of observables}, and one must define an equivalent to
quantum mechanics without Hilbert space \cite{Jac,Gr,G1,G2,Gr1}. Non-vanishing
3-cocycles arising in quantum field theory force us to go beyond the standard
approach and consider nonassociative generalization of the fibre bundle theory
and related nonassociative gauge theories \cite{N1,N2}. This work is in
progress.


\begin{thebibliography}{99}
\bibitem {Com} A failure of  the Jacobi identity also is known in quantum field theory
(see ref. [9] and references therein).
\end{thebibliography}
\end{document}